\documentclass[11pt,a4paper]{article}

\usepackage[T1]{fontenc}
\usepackage[utf8]{inputenc}
\usepackage{lmodern}
\usepackage{microtype}

\usepackage[a4paper,margin=2.5cm]{geometry}
\usepackage{setspace}
\usepackage{amsmath,amssymb,amsthm,mathtools}
\usepackage{bm}
\usepackage{siunitx}

\usepackage[dvipsnames,table]{xcolor}
\usepackage{graphicx}
\graphicspath{{figures/}}
\usepackage{booktabs}
\usepackage{multirow}
\usepackage{array}
\usepackage{placeins}

\usepackage{caption}
\usepackage{subcaption}

\usepackage{algorithm}
\usepackage{algpseudocode}

\usepackage[hidelinks,colorlinks=true,
            linkcolor=blue!60!black,citecolor=blue!60!black,
            urlcolor=blue!60!black]{hyperref}
\usepackage{orcidlink}
\usepackage[capitalise,noabbrev]{cleveref}

\usepackage[backend=bibtex,
            style=numeric-comp,
            sorting=none,
            giveninits=true,
            maxnames=6,minnames=1]{biblatex}
\usepackage{authblk}

\title{Reinforcement learning for vertical position control on the EXL-50U spherical tokamak}
\author[1,2,3]{Lei Xing~\orcidlink{0000-0001-7655-0920}\thanks{E-mail: \texttt{yfyg4669@outlook.com}}}
\author[1,2,3]{Huicong Ma}
\author[5]{Changquan Yu}
\author[1,2,3]{Xuanhe Wang}
\author[7]{Jiayi Zhi}
\author[4]{Pei Guo}
\author[6]{Mengyao Li}
\author[1,2,3]{Zhengyuan Chen}
\author[1,2,3]{Yapeng Zhang}
\author[1,2,3]{Guoyang Shi}
\author[1,2,3]{Dongkai Qi}
\author[1,2,3]{Xiang Gu}
\author[1,2,3]{Siqi Ding}
\author[1,2,3]{Yong Liu}
\author[1,2,3]{Jianguo Chen}
\author[1,2,3]{Tianyuan Liu$^{*}$}
\author{and the EXL-50U Team$^{a}$}

\affil[1]{Beijing ENN Fusion Energy Science and Technology Co., Ltd., Beijing 101111, China}
\affil[2]{Beijing Key Laboratory of High Magnetic Field Spherical Torus Fusion Energy}
\affil[3]{Hebei Key Laboratory of Compact Fusion}
\affil[4]{Key Laboratory of Materials Modification by Beams of the Ministry of Education, School of Physics, Dalian University of Technology, Dalian 116024, China}
\affil[5]{School of Nuclear Science and Engineering, East China University of Technology, Nanchang 330013, Jiangxi, China}
\affil[6]{Dalian University of Technology, Dalian 116024, China}
\affil[7]{Institute of Robotics and Automatic Information System, Nankai University, Tianjin 300350, China}
\affil[*]{Corresponding author: \texttt{tianyuanliu1992@gmail.com}}
\affil[a]{See Shi~\textit{et~al} 2025 (\url{https://doi.org/10.1088/2058-6272/ad9e8f}) for the EXL-50U Team.}

\date{\today}

\begin{document}

\maketitle

\begin{abstract}
Vertical position control is essential for sustaining high-performance operation in spherical tokamaks, 
where increased plasma elongation introduces stringent requirements on fast and robust stabilization. 
This work presents an experimentally validated reinforcement-learning (RL)-based vertical position control framework for the EXL-50U spherical tokamak.
A high-fidelity discharge-reconstructed simulation environment is developed by integrating physics-based plasma–circuit models with experimental equilibrium information, 
enabling systematic controller synthesis and sim-to-real evaluation. 
Within this framework, RL is benchmarked in simulation against operational
proportional--integral--derivative (PID) and model-based linear quadratic
regulator (LQR) controllers under identical plant dynamics, actuator
constraints, and measurement imperfections.
Simulation results show that RL achieves tracking accuracy comparable to PID
with consistently lower vertical-stabilization coil effort, while lightweight
integral compensation improves robustness against residual model--plant
mismatch.
The RL controller is subsequently deployed on EXL-50U for closed-loop experiments. 
Across more than ten discharges with RL takeover, stable vertical regulation is achieved within the controlled windows.
For seven representative discharges, RL maintains millimetre-scale tracking accuracy comparable to the operational PID controller (MAE typically ~1–5 mm) while consistently reducing actuator effort. 
These results demonstrate the feasibility of learning-based plasma control on a real spherical tokamak and establish a practical pathway toward future fusion control systems.

\end{abstract}


\section{Introduction}
\label{sec:introduction}

Controlled nuclear fusion is widely regarded as a long-term pathway to clean,
abundant, and low-carbon baseload energy, with magnetic confinement in tokamak
and related toroidal configurations remaining among the most mature routes
toward a practical power plant.
Within this landscape, ENN Science and Technology Development Co., Ltd.\ has
advanced a spherical-torus (ST) roadmap for proton--boron (p--$^{11}$B) fusion
as a compact, aneutronic path toward clean energy~\cite{Liu2024ENNRoadmap};
the upgraded device EXL-50U ($R_0\sim0.6$--$0.8\,\mathrm{m}$,
$A\sim1.4$--$1.8$, $B_T$ up to $1.2\,\mathrm{T}$) serves as the near-term
experimental platform supporting the next-step EHL-2
design~\cite{Shi2025EXL50UStrategy,Shi2026EXL50UOverview}, and recent campaigns
have already demonstrated high-parameter p--B operation, including the first
publicly reported MA-class hydrogen--boron discharges with $\sim10\%$ boron
fuel fraction~\cite{Shi2025EXL50U1MA} and the first H-mode in p--$^{11}$B
plasmas under NBI heating~\cite{Wang2026FirstHmodePB}.

Elongated tokamak plasmas require active stabilization of an axisymmetric
vertical instability~\cite{Lazarus1990VerticalControl}.
Spherical tori exploit natural elongation for compact high-performance
equilibria~\cite{Ono2015STReview}, so that as operations push toward higher
elongation the margin for vertical-position control becomes correspondingly
more critical~\cite{Berkery2023MASTUOps}---even though the underlying $n=0$
control problem remains homologous to that on conventional-aspect-ratio
devices.
On EXL-50U, the same high-parameter elongated plasmas that enable the
p--B programme also approach regimes where uncontrolled vertical
displacement events (VDEs) can impose severe electromagnetic loads and
threaten structural integrity, as quantified by disruption-oriented 3D EM
assessments~\cite{EXL50UDisruptionEM2026}.
Reliable vertical stability control is therefore a prerequisite for
sustaining and extending such high-performance operation.

\begin{figure*}[tbp]
  \centering
  \includegraphics[width=\textwidth]{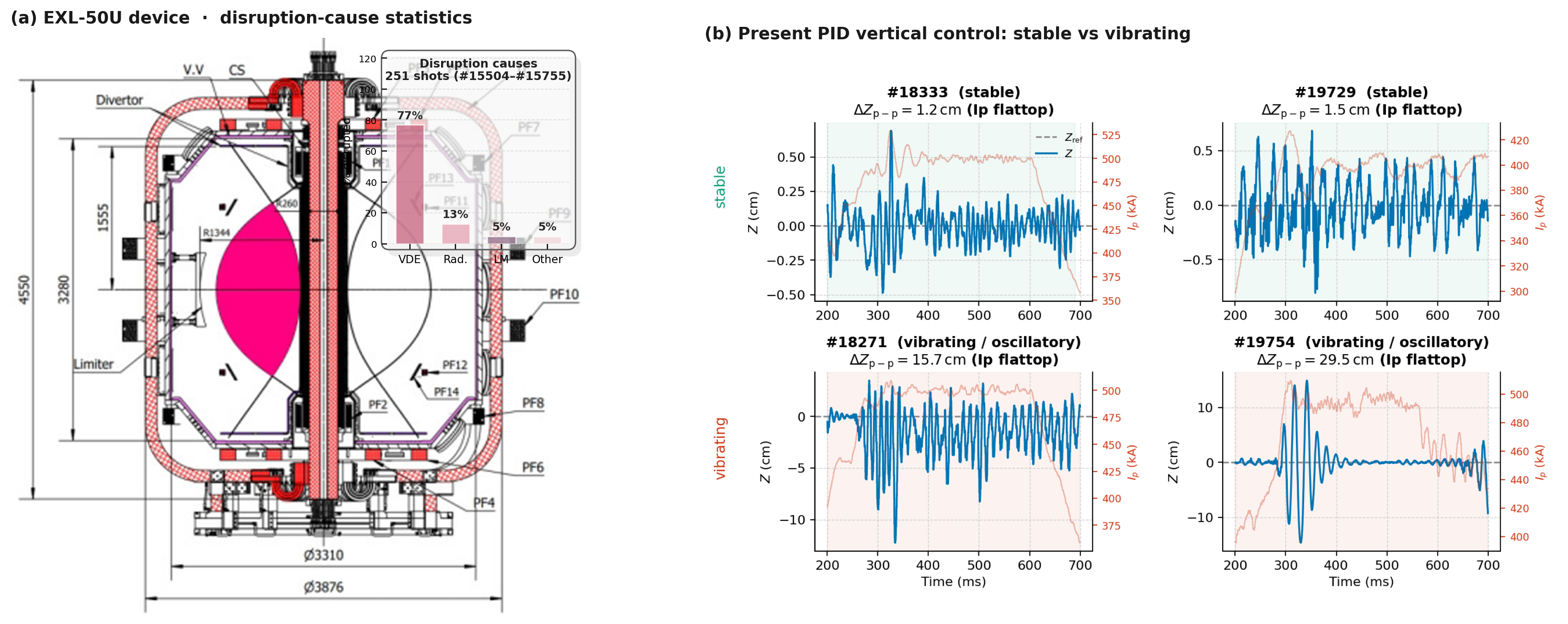}
  \caption{EXL-50U vertical-control motivation.
  (a)~Device cross-section with an inset bar chart of disruption causes for
  shots \#15504--\#15755 (251 discharges; 95 disrupted, rate $38\%$; among
  disrupted shots: VDE $77\%$, radiation $13\%$, locked mode $5\%$, others $5\%$).
  (b)~Examples of present PID vertical position control (200--700\,ms):
  top row, comparatively stable (\#18333, \#19729); bottom row, vibrating /
  oscillatory (\#18271, \#19754).
  Blue: measured $Z$; dashed grey: $Z_{\mathrm{ref}}$ (CCPV); light red: $I_p$.
  Peak-to-peak tracking error $\Delta Z_{\mathrm{p-p}}$ is evaluated on the
  $I_p$ flattop (shaded).}
  \label{fig:exl50u-vde-motivation}
\end{figure*}

Disruptions---and the VDE path in particular---are a shared operational
risk across tokamaks and spherical tori.
A decade-long survey of 2309 JET disruptions ($I_p>1\,\mathrm{MA}$)
documented how improved scenarios reduced the disruption rate from about
$15\%$ to below $4\%$~\cite{deVries2011JETDisruptionSurvey}, while a still
larger JET database of 4854 disruptions found that about $41\%$ developed
strong plasma-current asymmetries of the kind associated with VDEs, with
upward VDEs dominating that asymmetrical
subset~\cite{Gerasimov2014JETAsymDisruption}.
Multi-device collation under the ITPA Disruption Database likewise frames
VDE-linked electromagnetic loads as a common concern on both conventional
and spherical devices~\cite{Eidietis2015ITPAIDDB}.
Against this backdrop, a recent EXL-50U window (shots \#15504--\#15755;
251 discharges spanning p--B, H-mode, and NBI ion-temperature enhancement
experiments) recorded a disruption rate of $38\%$, with VDE accounting for
$77\%$ of the identified causes---a campaign-dependent share that, while
higher than the $\sim40\%$ asymmetrical/VDE-linked order seen in the JET
databases above, underscores the same common risk rather than an isolated
anomaly.
\Cref{fig:exl50u-vde-motivation}(a) summarizes the machine geometry
together with these disruption statistics.
Under the present PID-based vertical position control, limiter plasmas can
typically be held at the millimetre level, whereas divertor configurations
commonly retain residual vertical oscillations of order
$5\,\mathrm{cm}$ (\cref{fig:exl50u-vde-motivation}(b)).
Such large motion degrades discharge reliability and, through
antenna--plasma gap variation, also impairs ICRH
coupling~\cite{Zhang2024EASTICRF}.

How the community addresses this need can be compared across major tokamaks
and spherical tori along three practical axes: \emph{models},
\emph{algorithms}, and \emph{robustness measures} that keep a fast vertical
loop effective under changing equilibria, delays, and actuator limits.

On the modeling side, high-fidelity vertical plants are built from
electromagnetic descriptions that couple plasma vertical motion to vessel
eddy currents and control coils.
Linearized rigid-body or deformable plasma--circuit models---typified by
RZIP-type formulations on EAST, with closely related linear plants used on
TCV and in ITER vertical-stability design
studies~\cite{Rui2024EASTLQR,Pesamosca2022TCVHinf,Ambrosino2014ITERMagnetic,Ambrosino2009ITERVS}---provide
the synthesis and closed-loop evaluation models for most vertical
controllers.
When large displacements or nonlinear evolution must be resolved,
free-boundary nonlinear simulators are used instead---for example
\mbox{NSFsim} on DIII-D, which couples the free-boundary
Grad--Shafranov equation to one-dimensional core
transport~\cite{Subbotin2026DIIIDRL}.

On the algorithm side, classical PD/PID (or PD-like) feedback remains the
operational baseline on most devices, including
EAST~\cite{Rui2024EASTLQR}, MAST-U~\cite{MASTU_PCS_VS},
TCV~\cite{Pesamosca2022TCVHinf},
DIII-D~\cite{Schuster2005DIIIDAntiWindup}, and
Globus-M/M2~\cite{Konkov2020GlobusLMI}.
Beyond this baseline, model-based optimal, robust, and learning controllers
have been developed to improve gain selection, multi-coil coordination, and
closed-loop performance: LQR/LQG designs on EAST and in the ITER control
literature~\cite{Rui2024EASTLQR,Rui2026EASTAdaptive,Ambrosino2014ITERMagnetic,Ambrosino2009ITERVS};
structured $H_\infty$ and LMI-based robust designs on TCV, DIII-D,
and Globus-M/M2~\cite{Pesamosca2022TCVHinf,Humphreys2000DIIIDMIMO,Walker2003DIIIDHinf,Konkov2020GlobusLMI};
and deep reinforcement learning for magnetic / vertical control on TCV and
DIII-D~\cite{Degrave2022TCVRL,Tracey2024TCVRL,Subbotin2026DIIIDRL}.
These model-based approaches share a common premise: once a credible plant
model or training environment exists, controller parameters can be obtained
systematically rather than only by empirical scans.

Beyond the feedback law itself, devices also pursue complementary
robustness aids that keep a fast vertical loop effective as latencies and
equilibria change.
On the latency side, low-latency FPGA (or equivalent dedicated) loops
reduce control-cycle delay, as on MAST-U's P6
system~\cite{MASTU_PCS_VS}, while delay compensation---e.g., Smith
predictors with adaptive LQR on EAST---enlarges the controllable
growth-rate window~\cite{Rui2026EASTAdaptive}.
On the model-adaptation side, neural-network real-time response
identification retunes model-based gains during the
pulse~\cite{Rui2024EASTLQR,Rui2026EASTAdaptive}.

Taken together, the multi-device picture indicates that classical PD/PID
remains the operational baseline on many devices, while model-based and
learning-based feedback can set vertical-loop gains more systematically
once a credible plant model or training environment is available.
Such robustness aids are clearly important, but they are not the first
step taken here: this work instead develops a reinforcement-learning
policy as the primary vertical controller for EXL-50U---augmented by a
simple integral term for model--plant mismatch---while retaining classical
PID and LQR as engineering and model-based baselines for systematic
comparison in high-fidelity simulation and on-device experiments, leaving
stronger sim-to-real aids (e.g., Smith predictors or online identification)
to subsequent development.
\Cref{tab:vde-method-devices} summarizes representative method families by
device and situates the RL-centred approach considered for EXL-50U in this
work against the PID and LQR baselines.

\begin{table}[t]
  \centering
  \caption{Representative vertical-stability / position-control method families
  by device (portrait layout: devices as rows).
  Check marks denote publicly reported \emph{device} implementations
  (experiment or operations), each with a supporting citation; they are not
  claimed to be unique or currently the default PCS law on every device.
  Simulation-only studies are excluded.
  Learning-aided \emph{parameter identification} that retunes a classical /
  LQR law is cited under LQR/LQG (not under RL).
  An RL tick is granted when a deployed policy includes vertical position /
  location (or equivalent elongation / vertical-stability objectives) among
  its controlled targets or reward terms.
  The final row lists the methods considered for EXL-50U in this work:
  reinforcement learning as the primary controller, with engineering PID
  and model-based LQR retained as baselines, as evaluated in high-fidelity
  simulation and on-device experiments.}
  \label{tab:vde-method-devices}
  \small
  \setlength{\tabcolsep}{4.5pt}
  \begin{tabular}{@{}lcccc@{}}
    \toprule
    Device
      & PD/PID
      & LQR/LQG
      & $H_\infty$/LMI
      & RL \\
    \midrule
    EAST
      & \checkmark\,\cite{Rui2024EASTLQR}
      & \checkmark\,\cite{Rui2024EASTLQR,Rui2026EASTAdaptive}$^{\ddagger}$
      &
      &  \\
    MAST-U
      & \checkmark\,\cite{MASTU_PCS_VS}
      &
      &
      &  \\
    TCV
      & \checkmark\,\cite{Pesamosca2022TCVHinf}
      &
      & \checkmark\,\cite{Pesamosca2022TCVHinf}
      & \checkmark\,\cite{Degrave2022TCVRL,Tracey2024TCVRL}$^{\S}$ \\
    DIII-D
      & \checkmark\,\cite{Schuster2005DIIIDAntiWindup}$^{\ast}$
      &
      & \checkmark\,\cite{Humphreys2000DIIIDMIMO,Walker2003DIIIDHinf}
      & \checkmark\,\cite{Subbotin2026DIIIDRL} \\
    ITER$^{\dagger}$
      &
      & \checkmark\,\cite{Ambrosino2014ITERMagnetic,Ambrosino2009ITERVS}
      &
      &  \\
    Globus-M/M2
      & \checkmark\,\cite{Konkov2020GlobusLMI}$^{\ast\ast}$
      &
      & \checkmark\,\cite{Konkov2020GlobusLMI}
      &  \\
    \midrule
    \textbf{This work (EXL-50U)}
      & \checkmark
      & \checkmark
      &
      & \checkmark \\
    \bottomrule
  \end{tabular}

  \vspace{0.4em}
  {\footnotesize
  $^{\dagger}$Design / simulation studies.
  $^{\ddagger}$Includes neural-network real-time response-parameter
  identification that retunes adaptive LQR (not a standalone RL policy).
  $^{\S}$Integrated magnetic RL on TCV (experiment): vertical position /
  location among controlled objectives and reward terms; not a dedicated
  single-loop VS controller.
  $^{\ast}$Nominal vertical loop with anti-windup augmentation.
  $^{\ast\ast}$Cascade / MIMO PID designs under LMI constraints
  (listed under both PD/PID and $H_\infty$/LMI).}
\end{table}

Motivated by this EXL-50U operational gap and by multi-device evidence that
model-based and learning-based feedback can enlarge the controllable
operating space, this work establishes a vertical-position control
framework for EXL-50U in which a reinforcement-learning policy constitutes
the primary controller, while classical PID and LQR controllers are
retained as engineering and model-based baselines for systematic
comparison, with integral compensation introduced to mitigate
model--plant / training--deployment mismatch (state augmentation for LQR;
an external integral for RL).
A high-fidelity, discharge-reconstructed simulation environment is used
for synthesis and training and for extensive ablation---including
measurement-chain imperfections, actuator constraints, and integral-action
variants---to quantify model--plant mismatch and reduce the sim-to-real
gap prior to deployment.
On-device closed-loop tests further delineate the attainable performance of
each method: across more than ten discharges with RL takeover, the policy
maintained stable vertical-position regulation during the controlled
window, clarifying gains and limits relative to the PID and LQR baselines
in recovery speed, tracking accuracy, and actuation cost.

The main contributions of this work are threefold:
\begin{enumerate}
  \item A high-fidelity discharge-reconstructed simulation environment is
        established for vertical control synthesis and sim-to-real
        evaluation.
  \item PID, LQR, and reinforcement-learning controllers are systematically
        compared in simulation under identical plant dynamics and actuator
        constraints.
  \item Reinforcement-learning-based vertical position control is
        experimentally demonstrated on EXL-50U with stable closed-loop
        regulation over multiple discharges.
\end{enumerate}

The remainder of this paper is organized as follows.
\Cref{sec:control_methods} presents the control-design framework for EXL-50U
along the device--model--controller--sim2real stack: the plant and control
objective, the PID (model-free) and LQR (model-based) baselines, the
reinforcement-learning policy as the primary controller, and integral
compensation (state-augmented for LQR, external for RL) as the concrete entry
in a coarse mismatch-remedy layer.
\Cref{sec:simulation} describes the high-fidelity, discharge-reconstructed
simulation environment, including integral-action ablation and comparisons
relative to the baselines.
\Cref{sec:experiment} reports on-device closed-loop tests, with representative
RL and LQR pulses and a multi-shot RL takeover campaign.
\Cref{sec:conclusion} summarizes the main findings and outlines directions
for stronger sim-to-real compensation.

\section{Control Methods for Vertical Stability}
\label{sec:control_methods}

\begin{figure}[htbp]
  \centering
  \includegraphics[width=0.70\linewidth]{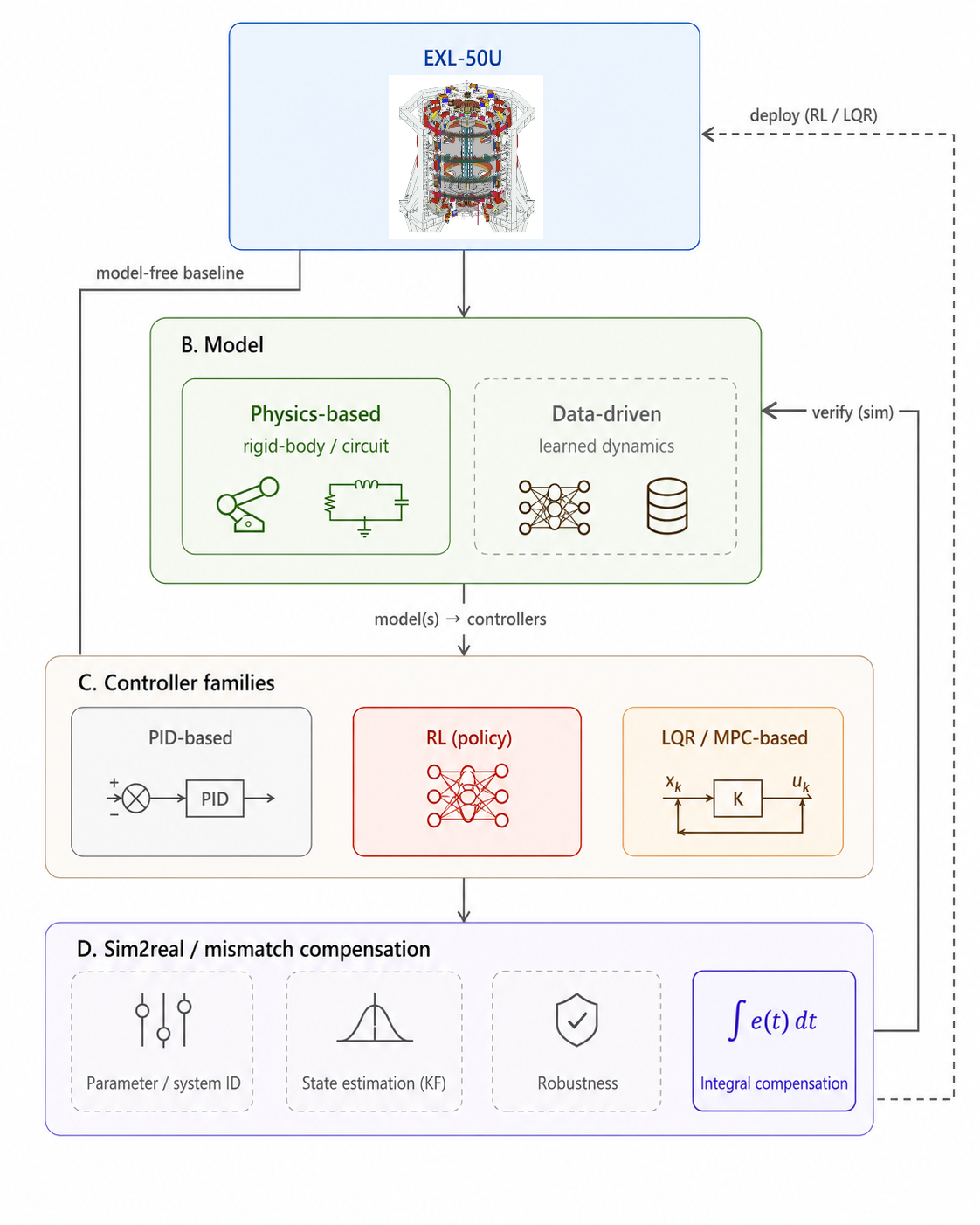}
  \caption{Control-design framework for vertical position control on EXL-50U.
  (A)~Device experiment: platform data and the operational PID baseline
  (model-free path into~C).
  (B)~Model layer: physics-based plant (rigid-body / circuit; used in this
  work) and a data-driven slot (placeholder; not developed here);
  model(s) feed the controllers.
  (C)~Controller families: PID-based (engineering baseline), RL policy
  (primary), and LQR~/~MPC-based (model-based family; LQR delivered here).
  (D)~Sim2real~/~mismatch remedies (coarse): integral compensation
  (this work; LQR$\to$LQRI by state augmentation, RL$\to$RLI by an external
  integral), with system identification, state estimation, and robust
  synthesis as outlook items.
  Closed loops: verify in simulation and deploy RL~/~LQR to the device.}
  \label{fig:control-framework}
\end{figure}

This section presents the vertical-position control framework developed for
EXL-50U.
\Cref{fig:control-framework} summarizes the four-layer workflow.
Device experiment~(A) supplies platform data and the operational PID baseline.
The model layer~(B) is headed by a physics-based plant used throughout this
work, with a data-driven slot shown only as a placeholder; models feed the
controller families~(C).
Among those families, classical PID is retained as the model-free engineering
baseline, LQR as the delivered member of the LQR~/~MPC-based family, and a
reinforcement-learning policy as the primary controller.
A separate sim2real~/~mismatch layer~(D) then hosts coarse remedies: this work
implements integral compensation on the LQR and RL paths (LQR$\to$LQRI by
state augmentation; RL$\to$RLI by an external integral), while system
identification, state estimation, and robust synthesis remain outlook items.
Verification in simulation and on-device deployment of RL~/~LQR close the loop
back to~(B) and~(A).
The subsections below follow this hierarchy: plant and control objective,
baseline laws (PID, LQR), the RL design, and finally the integral augmentation
as the concrete D-layer add-on.
Sampling, measurement imperfections, actuator limits, and related
simulation-environment issues are deferred to~\cref{sec:simulation}.

\subsection{Plant model and control objective}
\label{sec:plant_model}

High plasma elongation improves tokamak confinement but is associated with a negative decay index,
\begin{equation}
n=-(R/B_z)\partial B_z/\partial R<0,
\end{equation}
of the vertical equilibrium field, which makes the plasma vertically unstable: any vertical displacement grows exponentially in open loop.

On the control timescale the plasma is massless and in instantaneous force balance, governed by the Grad--Shafranov (GS) equation:
\begin{equation}
\Delta^*\psi=-2\pi\mu_0 R\,(j_{pl,\phi}+j_{e,\phi}),\qquad
\Delta^*=R\frac{\partial}{\partial R}\!\left(\frac{1}{R}\frac{\partial}{\partial R}\right)+\frac{\partial^2}{\partial Z^2},
\end{equation}
where \(j_{pl,\phi}\) is set by the flux functions \(p(\psi)\) and \(F(\psi)=RB_\phi\), and \(j_{e,\phi}\) is the toroidal current density in the external conductors (coils, passive plates, vessel). The rigid-plasma RZIP model approximates the plasma current distribution as a rigid translation of its equilibrium distribution: mass is neglected and the plasma moves rigidly, retaining only the radial and vertical positions of the current centroid and the total plasma current, with circuit coupling among active coils, passive plates, and vacuum vessel.

In the full circuit model, conductors are coupled by geometric mutual inductances. Faraday's laws for the active coils and plasma circuit are
\begin{align}
M_{cc}\dot{I}_c+M_{c,\mathrm{pass}}\dot{I}_{\mathrm{pass}}+M_{cv}\dot{I}_v+\dot{\Psi}_{cp}+R_c I_c&=V_c,\\
\dot{\Psi}_{pp}+R_p I_p+M_{pc}\dot{I}_c+M_{p,\mathrm{pass}}\dot{I}_{\mathrm{pass}}+M_{pv}\dot{I}_v&=0,
\end{align}
where \(I_c\), \(I_{\mathrm{pass}}\), \(I_v\), \(I_p\) are the active-coil, passive-plate, vacuum-vessel, and plasma currents; \(M_{ab}\) is a mutual inductance; \(R_c\) and \(R_p\) are the coil and plasma resistances; \(V_c\) is the applied voltage; and \(\Psi_{cp}\) is the plasma-generated flux at the coils. Analogous equations hold for the passive plates and vessel, with zero applied voltage. These equations are nonlinear in the currents because the plasma flux depends on the current distribution and position.

Linearizing about a nominal equilibrium (subscript ``eq''), with perturbations \(\delta I=I-I_{\mathrm{eq}}\), \(\delta y=y-y_{\mathrm{eq}}\), \(\delta Z=Z-Z_{\mathrm{eq}}\), the current centroid \((R_{\mathrm{cur}},Z_{\mathrm{cur}})\) follows the conductor currents. Instantaneous force balance linearizes to
\begin{align}
\frac{\partial F_r}{\partial R}\,\delta R+\frac{\partial F_r}{\partial I}\,\delta I&=0,\\
\frac{\partial F_z}{\partial Z}\,\delta Z+\frac{\partial F_z}{\partial I}\,\delta I&=0,
\end{align}
which gives the centroid sensitivities
\begin{equation}
\frac{\partial R_{\mathrm{cur}}}{\partial I}=-\Bigl(\frac{\partial F_r}{\partial R}\Bigr)^{-1}\frac{\partial F_r}{\partial I},\qquad
\frac{\partial Z_{\mathrm{cur}}}{\partial I}=-\Bigl(\frac{\partial F_z}{\partial Z}\Bigr)^{-1}\frac{\partial F_z}{\partial I}.
\end{equation}
Substituting these into the position derivatives of the plasma flux yields the plasma response matrices \(X_{ab}\), giving the plasma-modified inductance \(M_{ab}^{*}=M_{ab}+X_{ab}(J,B_{\mathrm{vac}})\), with \(B_{\mathrm{vac}}\) the vacuum field at the linearization point. The motion-related part of \(\Psi_{cp}\) is absorbed into \(X_{ab}\). Assembling the circuit blocks gives the compact conductor--plasma circuit equation
\begin{equation}
M^{*}\dot{I}+RI=\Gamma V,
\end{equation}
where \(M^{*}\) is the plasma-modified inductance, \(R\) is the resistance, \(\Gamma\) maps applied voltages to circuit ports, and \(V\) is the actual applied voltage.

The assembled plant may be written in nonlinear state-space form \(\dot{\mathbf{x}}=f(\mathbf{x},\mathbf{u})\), \(\mathbf{y}=h(\mathbf{x},\mathbf{u})\) with state \(\mathbf{x}=[I_c;I_{\mathrm{pass}};I_v;I_p]\). Here \(\mathbf{u}\) denotes the voltage command, while the plant input is the applied voltage \(\mathbf{V}\) (\(\mathbf{V}=\mathbf{a}\odot\mathbf{u}+\mathbf{b}\), with channel gain \(\mathbf{a}\) and bias \(\mathbf{b}\)). For linear controller synthesis the model is linearized about an equilibrium \((\mathbf{x}_{\mathrm{eq}},\mathbf{V}_{\mathrm{eq}})\). The linear state-space plant is
\begin{equation}
\delta\dot{\mathbf{x}}=A\,\delta\mathbf{x}+B\,\delta\mathbf{V},\qquad
\delta\mathbf{y}=C\,\delta\mathbf{x}+D\,\delta\mathbf{V},
\end{equation}
where \(A\), \(B\), \(C\), and \(D\) represent the linearized state, input, output, and feedthrough matrices (\(D=0\)), respectively.

Fast vertical stabilization (VS \(\rightarrow Z\)) is treated separately from the slower radial-position and shape-control loops, since they operate on different timescales; the radial degree of freedom is retained only for electromagnetic consistency. The control objective is to stabilize the open-loop unstable mode with growth rate \(\gamma_z=\max_i\operatorname{Re}\lambda_i(A)\), using \(u_{vs}\) so that
\begin{equation}
Z_{\mathrm{cur}}(t)\to Z_{\mathrm{ref}}(t),
\end{equation}
where \(Z_{\mathrm{ref}}(t)\) denotes the desired plasma vertical position. During operation, the applied voltage is restricted by the available power supply capability,
\begin{equation}
V_{vs,\min}\le V_{vs}\le V_{vs,\max}.
\end{equation}
All controllers in this section are evaluated against the same plant family and
the same actuator limits, so that differences in closed-loop behaviour can be
attributed to the feedback law rather than to inconsistent modelling
assumptions.
Within that shared setting, the model layer of~\cref{fig:control-framework}
provides the representations consumed by the controllers: LQR synthesis uses a
reduced-order VS$\to Z$ channel, while RL training and closed-loop verification
use the full-order plant; fidelity choices are detailed in the LQR and RL
subsections and are not drawn as separate branches on the framework figure.
The PID baseline is model-free and does not consume either representation for
gain design.
Model-reduction checks and practical closed-loop imperfections are deferred to
the LQR design and to~\cref{sec:simulation}, respectively.

\subsection{PID baseline (model-free)}
\label{sec:pid-baseline}

An incremental PID law is retained as the engineering / model-free baseline
already used in EXL-50U operations (\cref{fig:control-framework}).
It is not proposed as a new synthesis method here; rather, it provides the
reference against which the model-based controllers are compared.
The controller updates the VS-coil voltage from the tracking error between the
reference and measured vertical positions. At the sampling instant $k$, the
position error is defined as

\begin{equation}
e_k=Z_{\mathrm{ref}}-Z_{\mathrm{cur},k}.
\end{equation}

The incremental control law is given by

\begin{equation}
\Delta u_{vs,k}
=
K_p(e_k-e_{k-1})
+
K_i e_k\Delta t
+
\frac{K_d}{\Delta t}
(e_k-2e_{k-1}+e_{k-2}),
\end{equation}

where $K_p$, $K_i$, and $K_d$ denote the proportional, integral, and derivative
gains. The integral and derivative gains are determined from the corresponding
time constants,

\begin{equation}
K_i=\frac{K_p}{T_i},
\qquad
K_d=K_pT_d .
\end{equation}

The VS coil voltage command is updated according to

\begin{equation}
u_{vs,k}
=
u_{vs,k-1}
-
\Delta u_{vs,k},
\end{equation}

where the sign convention is selected according to the polarity relationship
between the VS coil magnetic field and the plasma vertical displacement.
The final command is saturated within the actuator limits
\(V_{vs,\min}\le u_{vs,k}\le V_{vs,\max}\).

\subsection{LQR baseline (model-based)}
\label{sec:lqr-baseline}

A linear quadratic regulator (LQR) is retained as the model-based baseline
(\cref{fig:control-framework}): gains are synthesized from a reduced plant and
serve as a classical optimal-feedback reference for comparison with the RL
policy.
Unlike the model-free PID law, LQR uses an explicit linear response model and
penalizes both tracking error and actuation effort.

\begin{figure}[htbp]
  \centering
  \includegraphics[width=0.98\linewidth]{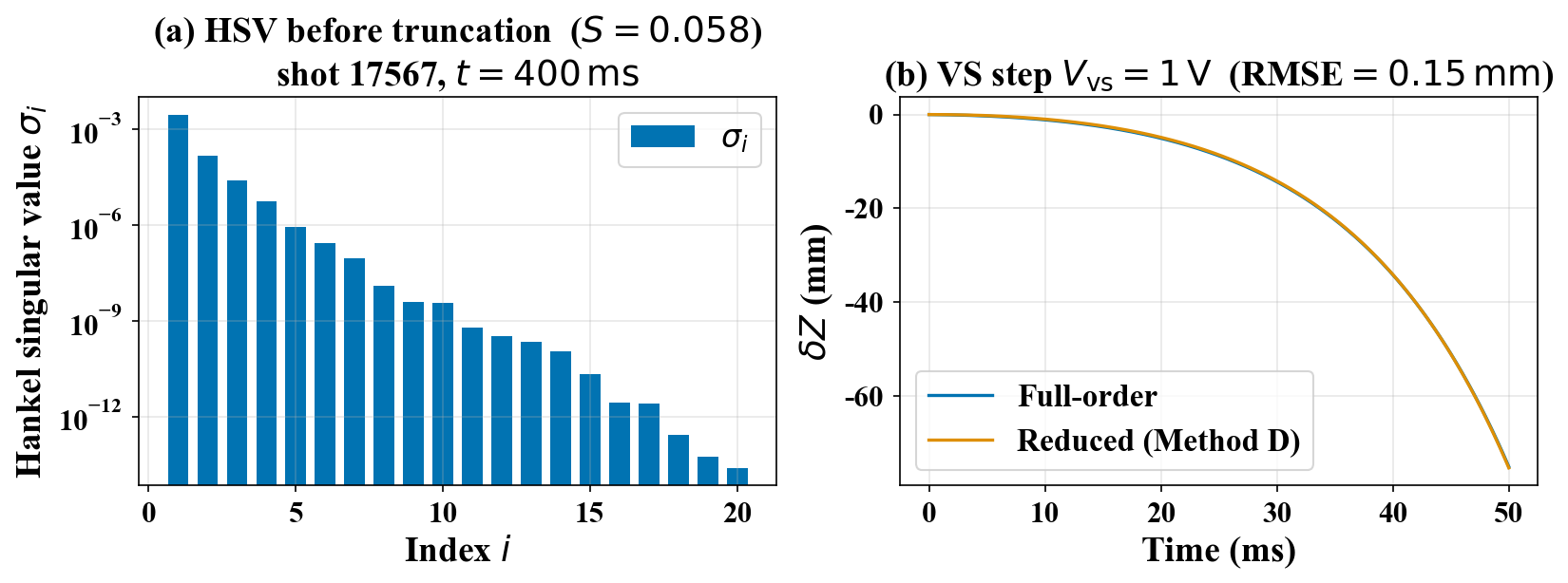}
  \caption{Adequacy check of the reduced plant used by LQR
  (shot~\#17567, \(t=400\,\mathrm{ms}\)), in the style of~\cite{Rui2024EASTLQR}.
  (a)~Hankel singular values of the stable subsystem and index \(S\).
  (b)~Open-loop VS step responses (\(V_{\mathrm{vs}}=1\,\mathrm{V}\)):
  full-order plant versus the two-state model used for LQR synthesis.}
  \label{fig:full-vs-reduced}
\end{figure}

Dominant VS$\to Z$ dynamics are represented by the two-state reduced model
\begin{equation}
\begin{gathered}
\delta\dot{\mathbf{x}}_z
=
\mathbf{A}_z\delta\mathbf{x}_z
+
\mathbf{B}_z u_{vs},
\qquad
\delta Z_{\mathrm{cur}}
=
\mathbf{C}_z\delta\mathbf{x}_z ,
\end{gathered}
\end{equation}
with
\(\delta\mathbf{x}_z=[\delta Z_{\mathrm{cur}},\,\delta\dot Z_{\mathrm{cur}}]^T\).
The matrices are obtained by reducing the full-order VS$\to Z$ channel,
following the EAST LQR route~\cite{Rui2024EASTLQR}: Schur isolation of the
unstable growth rate \(P_1=\gamma\), balanced truncation of the stable
subsystem to a single pole \(P_2<0\), and a least-squares fit of the step gain
\(k\), giving
\begin{equation}
G(s)
=
\frac{k}{(s-P_{1})(s-P_{2})}
\;\Longleftrightarrow\;
\mathbf{A}_{z}
=
\begin{bmatrix}
0 & 1 \\
-P_{1}P_{2} & P_{1}+P_{2}
\end{bmatrix},
\quad
\mathbf{B}_{z}
=
\begin{bmatrix}
0 \\ k
\end{bmatrix},
\quad
\mathbf{C}_{z}
=
\begin{bmatrix}
1 & 0
\end{bmatrix}.
\end{equation}
Adequacy of this reduced plant is checked in~\cref{fig:full-vs-reduced}
(truncated-energy fraction
\(S=\sum_{i\ge 2}\sigma_{i}/\sum_{i}\sigma_{i}\approx 0.058\) on
shot~\#17567 at \(t=400\,\mathrm{ms}\); full-order dimension \(n=386\)).

On this reduced plant, infinite-horizon LQR minimizes
\begin{equation}
J
=
\int_0^\infty
\bigl(
\delta\mathbf{x}_z^{T}\mathbf{Q}\,\delta\mathbf{x}_z
+
R_u\,u_{vs}^{2}
\bigr)\,dt
\end{equation}
and yields the state-feedback law \(u_{vs}=-\mathbf{K}\,\delta\mathbf{x}_z\)
from the algebraic Riccati equation.
Integral augmentation for model--plant mismatch (LQRI) is deferred
to~\cref{sec:integral-comp}; it is not part of this baseline definition.

\subsection{Reinforcement learning policy (primary)}
\label{sec:rl-primary}

The primary controller developed in this work is a reinforcement-learning (RL)
policy for vertical position regulation
(\cref{fig:control-framework}).
In contrast to LQR, which synthesizes gains on a reduced VS$\to Z$ channel, the
policy is trained by interacting with the \emph{full-order} high-fidelity plant
introduced in~\cref{sec:plant_model}, so that actuator limits and the complete
circuit dynamics are present during learning.
PID and LQR remain the engineering and model-based baselines for comparison.

\paragraph{Task mapping.}
The vertical-control problem is cast as a Markov decision process
(MDP)~\cite{Sutton1998RL} with the following identification.
The environment is the full-order plasma--VS circuit model under the same
voltage limits \(V_{vs,\min}\le V_{vs}\le V_{vs,\max}\) used by the baselines.
At step \(t\), the agent observes the vertical displacement \(Z_t\) (relative
to the midplane target \(Z_{\mathrm{ref}}=0\)) and selects a normalized action
\(a_t\in[-1,1]\), which is mapped affinely to the physical VS voltage,
\begin{equation}
V_t
=
V_{\min}
+
\frac{1+a_t}{2}\,(V_{\max}-V_{\min}),
\end{equation}
with \(V_{\min}=-180\,\mathrm{V}\) and \(V_{\max}=180\,\mathrm{V}\) in the
present study.
The reward encourages reduction of the absolute vertical error,
\begin{equation}
r_t = |Z_{t-1}| - |Z_t|,
\label{eq:reward_z}
\end{equation}
and an episode terminates early if \(|Z_t|\ge 0.05\,\mathrm{m}\), or after a
fixed horizon (500 steps in training).
The learning objective is to maximize the expected discounted return
\begin{equation}
J(\theta)=\mathbb{E}_{\pi_\theta}\!\left[\sum_{t=0}^{T-1}\gamma^t r_t\right],
\end{equation}
where \(\pi_\theta(a_t\mid s_t)\) is the parameterized policy,
\(\gamma\in[0,1]\) the discount factor, and \(T\) the episode horizon.

\paragraph{Training algorithm.}
Policies are trained with Proximal Policy Optimization
(PPO)~\cite{Schulman2017PPO}, an on-policy actor--critic method whose clipped
surrogate limits overly large policy updates. With probability ratio
\(\rho_t(\theta)=\pi_\theta(a_t\mid s_t)/\pi_{\theta_{\mathrm{old}}}(a_t\mid s_t)\)
and advantage estimate \(\widehat A_t\), PPO maximizes
\begin{equation}
    L^{\mathrm{CLIP}}(\theta)
    = \mathbb{E}_t\!\left[
        \min\!\left(
            \rho_t(\theta)\,\widehat A_t,\;
            \operatorname{clip}\!\bigl(
                \rho_t(\theta), 1-\epsilon, 1+\epsilon\bigr)\,\widehat A_t
        \right)
    \right],
\end{equation}
augmented in the usual way by a value-function loss and an entropy bonus.
Network widths, learning-rate schedule, and related hyperparameters are
reported with the closed-loop simulation setup
in~\cref{sec:simulation}.
Integral compensation for residual model--plant / training--deployment
mismatch is introduced in~\cref{sec:integral-comp} as an optional add-on
(RL$\to$RLI), parallel to LQR$\to$LQRI, with different realizations on each
path.

\subsection{Integral compensation for model--plant mismatch}
\label{sec:integral-comp}

The PID, LQR, and RL designs above already define complete feedback laws.
In practice, however, reduced-model synthesis, training--deployment
differences, and slow drifts can leave a residual steady-state vertical
offset even when the loop is stabilizing.
Rather than introducing a fourth controller family, we place simple integral
compensation in the sim2real~/~mismatch layer~(D) of~\cref{fig:control-framework}
for both the LQR and RL paths---same role in the framework, different
mechanics---alongside outlook items such as system identification, state
estimation, and robust synthesis.

Define the accumulated tracking error
\begin{equation}
\eta(t)
=
\int_0^t
\bigl(Z_{\mathrm{ref}}(\tau)-Z_{\mathrm{cur}}(\tau)\bigr)\,d\tau .
\end{equation}
For LQR, \(\eta\) augments the reduced state to
\(\mathbf{x}_a=[\delta Z_{\mathrm{cur}},\,\delta\dot Z_{\mathrm{cur}},\,\eta]^T\),
and the Riccati design is repeated on the augmented plant to obtain LQRI,
\(u_{vs}=-\mathbf{K}_a\mathbf{x}_a\).
For RL, the same \(\eta\) is attached externally as an additional observation
(and/or a light additive correction on the policy voltage), yielding
RLI---by analogy with LQRI, the RL controller with external integral
compensation---used in ablation studies.
In both cases the integral term is a low-cost D-layer remedy---its necessity
and effect size are quantified in~\cref{sec:simulation}---not a standalone
algorithmic contribution.

\section{Simulation Study}
\label{sec:simulation}

This section develops and exercises the high-fidelity, discharge-reconstructed
simulation environment that underpins the control framework
of~\cref{sec:control_methods}.
The narrative follows three steps.
First, the reconstructed plant is verified against EXL-50U open- and
closed-loop measurements, and the practical non-idealities retained for all
subsequent tests (actuator limits, measurement chain) are stated once.
Second, on a fixed $+50\,\mathrm{mm}$ ramp--hold--return vertical reference,
we compare the engineering baseline (filtered PID), the model-based baseline
(LQR), and the primary RL policy---designed or trained on shot~\#17567 but
evaluated on shot~\#16235---under identical plant and noise conditions,
thereby stressing cross-shot transfer.
Third, the same reference profile is reused to ablate integral compensation
(LQR$\to$LQRI, RL$\to$RLI), quantifying its role against model--plant
mismatch and steady-state tracking offset.
A fixed-policy RLI check on four reconstructed plants then confirms
cross-shot tracking under the same reference; on-device multi-shot behaviour
is reported in~\cref{sec:experiment}.
Sampling, gains, and training hyperparameters follow~\cref{sec:control_methods}
unless restated below.

\subsection{Simulation environment verification}
\label{sec:sim-env-verify}

Before controller comparisons, the discharge-reconstructed simulation
environment is verified in \emph{closed loop} against EXL-50U measurements.
The plant is treated as a linear time-varying (LTV) state-space model: every
$10\,\mathrm{ms}$, the system matrices $(A,B)$ are replaced by a new
state-space model (SSM) computed from the experimental equilibrium and response
data of that shot, so that the simulated vertical dynamics track the evolving
operating point as closely as practicable, rather than freezing a single
linearization for the whole window.
Between updates, the discrete plant is integrated at the control sampling rate
$10\,\mathrm{kHz}$ ($\Delta t=0.1\,\mathrm{ms}$); the VS power supply is
modeled at $2\,\mathrm{kHz}$.
Verification is performed over $t\in[500,700]\,\mathrm{ms}$ on two
discharges spanning milder and stronger vertical growth,
\#16911 and \#17567 (blue: experiment; red: simulation).

Closed-loop tests use an incremental PID identical to the experimental law:
$K_p=2000$, $T_i=4\times10^{-4}\,\mathrm{s}$, $T_d=5\times10^{-3}\,\mathrm{s}$,
$|V_{\mathrm{VS}}|\le190\,\mathrm{V}$, and $Z_{\mathrm{ref}}=0$.
The VS command is computed online from the simulated $Z$ (\texttt{cmd = PID}).
The observed position is formed as $Z=k_Z C_Z I$, where the Z-gain $k_Z$
scales only the corresponding row of $C$ to correct diagnostic/reconstruction
mismatch and does not alter the PID gains or the scheduled $(A,B)$.
The values $k_Z=2.6$ and $k_Z=2.3$ are used for \#16911 and \#17567,
respectively.
Colored AR(1) noise is added to the $Z$ feedback channel, with
$(\sigma,\phi)=(3.5\times10^{-4},0.76)$ for \#16911 and
$(6.4\times10^{-4},0.8)$ for \#17567.

\begin{figure}[htbp]
\centering
\includegraphics[width=0.72\linewidth]{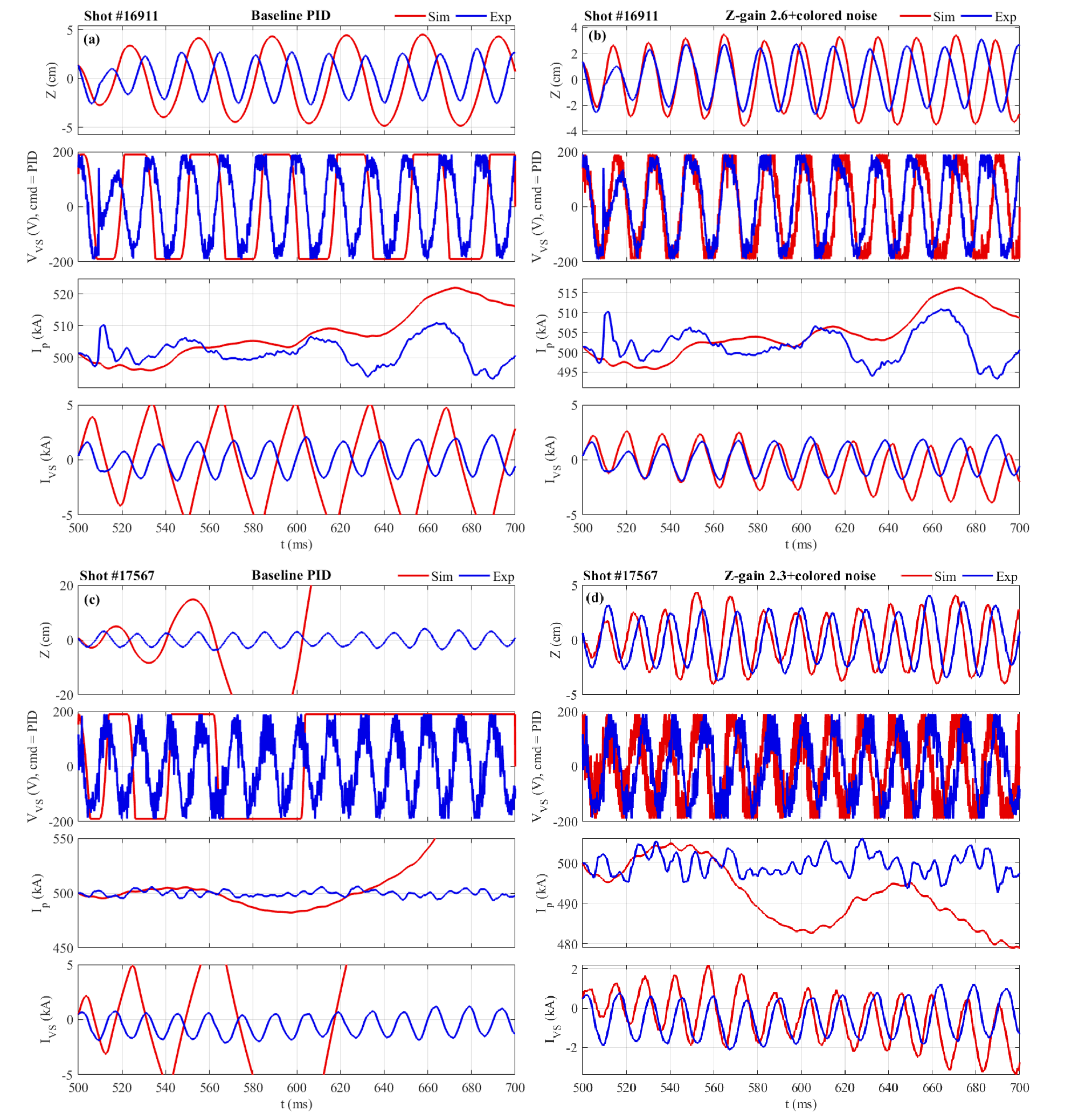}
\caption{PID closed-loop verification ($500$--$700\,\mathrm{ms}$) on the LTV
plant.
Top: \#16911 --- (a)~baseline; (b)~$k_Z=2.6$ + colored noise
($\sigma=3.5\times10^{-4}$, $\phi=0.76$).
Bottom: \#17567 --- (c)~baseline; (d)~$k_Z=2.3$ + colored noise
($\sigma=6.4\times10^{-4}$, $\phi=0.8$).
Blue: experiment; red: simulation.}
\label{fig:closed_pid_verify}
\end{figure}

For each shot, (a)~baseline PID ($k_Z=1$, no noise) is compared with (b)~the
shot-specific $k_Z$ and colored noise (\cref{fig:closed_pid_verify}).
On \#16911, setting~(b) improves the agreement in $Z$ and $I_{\mathrm{VS}}$ and
increases the high-frequency content of $V_{\mathrm{VS}}$ toward the experiment.
On \#17567, setting~(a) tends to large excursions or loss of control, whereas
setting~(b) maintains a bounded oscillation comparable to the experiment.
These closed-loop checks support using the LTV reconstructed plant for the
controller studies below.

Subsequent closed-loop comparisons retain the same shared non-idealities for
all controllers: a hard VS voltage bound
\(|V_{\mathrm{vs}}|\le 190\,\mathrm{V}\); an AR(1) measurement chain
(scale \(1.2\), bias \(-0.02\), \(\rho=0.76\),
\(\sigma=4.8\times10^{-4}\), delay \(1\,\mathrm{ms}\)) with a common seed per
shot; and a fixed plant/controller sign convention (polarity is not retuned
per method).

\subsection{Simulation setup}
\label{sec:sim-setup}

\paragraph{Plant and window.}
Controller comparisons use the same LTV reconstructed plant and shared
non-idealities as in~\cref{sec:sim-env-verify}, with control sampling
$\Delta t=0.1\,\mathrm{ms}$.
The model-based and learning controllers are synthesized / trained on
shot~\#17567; unless noted otherwise, the representative \emph{evaluation}
scenario is instead shot~\#16235 over a $T=250\,\mathrm{ms}$ window
(absolute times $t\in[400,650]\,\mathrm{ms}$), so that the tracking
comparisons probe cross-shot transfer rather than same-shot replay.
A short leading pad associated with measurement delay ($\approx 1\,\mathrm{ms}$)
is trimmed before scoring.

\paragraph{Reference profile.}
All tracking comparisons share one fixed vertical reference: relative to the
evaluation window, $Z_{\mathrm{ref}}$ starts at $0$, ramps linearly to
$+50\,\mathrm{mm}$, holds a plateau, then ramps linearly back to $0$ by the
end of the $250\,\mathrm{ms}$ window (ramp--hold--return; same shape as in
the \#17567 trajectory template, applied here on the \#16235 plant).
Steps~2--3 of this section (baseline comparison and integral ablation)
reuse this profile unchanged so that differences are attributable to the
controller, not to the reference schedule.

\paragraph{Controllers.}
Three designs from~\cref{sec:control_methods} enter the baseline comparison,
with gains held fixed (no per-shot retuning on \#16235):
\begin{itemize}
  \item \textbf{PID} (engineering baseline):
        $K_p=2000$, $T_i=5\,\mathrm{ms}$, $T_d=5\,\mathrm{ms}$, with a
        first-order derivative filter $\tau_{\mathrm{d}}=1\,\mathrm{ms}$
        (used throughout; no unfiltered PID variant is reported).
  \item \textbf{LQR} (model-based baseline):
        reduced-plant Riccati design of~\cref{sec:lqr-baseline} with
        $Q=\mathrm{diag}(10,\,10^{-3})$ and $R_u=10^{-5}$ (from \#17567).
  \item \textbf{RL} (primary): PPO policy of~\cref{sec:rl-primary}, trained on
        \#17567 for $10^{5}$ environment steps with maximum episode length
        $500$ steps.
        Key hyperparameters:
        $n_{\mathrm{steps}}=1024$, batch size $128$,
        network \([256,256,256,256]\), and reward
        \(r_t=|Z_{t-1}|-|Z_t|\) as in~\cref{eq:reward_z}.
\end{itemize}
Integral-augmented variants (LQRI, RLI) use the path-specific compensation
of~\cref{sec:integral-comp} with
$Q=\mathrm{diag}(10,\,10^{-3},\,3\times10^{5})$ for LQRI and are introduced
only in the integral-ablation comparison.

\paragraph{Performance indices.}
Tracking error is $e(t)=Z_{\mathrm{ref}}(t)-Z(t)$.
Four classical indices separate settling, peak deviation, accumulated error,
and actuation cost~\cite{SchultzRideout1961,GrahamLathrop1953}:
\begin{itemize}
  \item \textbf{Settling time} $t_s$: first instant after a short startup
        trim ($t_0=10\,\mathrm{ms}$) at which
        $|e(\tau)|\le\delta=5\,\mathrm{mm}$ for a contiguous dwell
        $\tau_{\mathrm{h}}=20\,\mathrm{ms}$,
        \begin{equation}
          t_s
          =\min\Bigl\{
            t\ge t_0:
            \bigl|e(\tau)\bigr|\le\delta
            \;\;\forall\,\tau\in\bigl[t,\,t+\tau_{\mathrm{h}}\bigr]
          \Bigr\}.
          \label{eq:kpi-ts}
        \end{equation}
        The same rule is used for all tracking tables in this section
        (including the four-shot RLI check).
  \item \textbf{Overshoot} $\mathrm{OS}$: peak of $|e(t)|$ over the window
        (equivalently, the largest absolute tracking deviation from
        $Z_{\mathrm{ref}}$).
  \item \textbf{ITAE:}
        \begin{equation}
          \mathrm{ITAE}
          =\int_{0}^{T} t\,\bigl|e(t)\bigr|\,\mathrm{d}t .
          \label{eq:kpi-itae}
        \end{equation}
  \item \textbf{Actuator effort:}
        \begin{equation}
          V_{\mathrm{rms}}
          =\sqrt{\frac{1}{T}\int_{0}^{T} V_{\mathrm{vs}}(t)^{2}\,\mathrm{d}t}.
          \label{eq:kpi-vrms}
        \end{equation}
\end{itemize}

\subsection{Baseline comparison: PID / LQR / RL}
\label{sec:sim-compare}

\begin{figure}[htbp]
  \centering
  \includegraphics[width=0.92\linewidth]{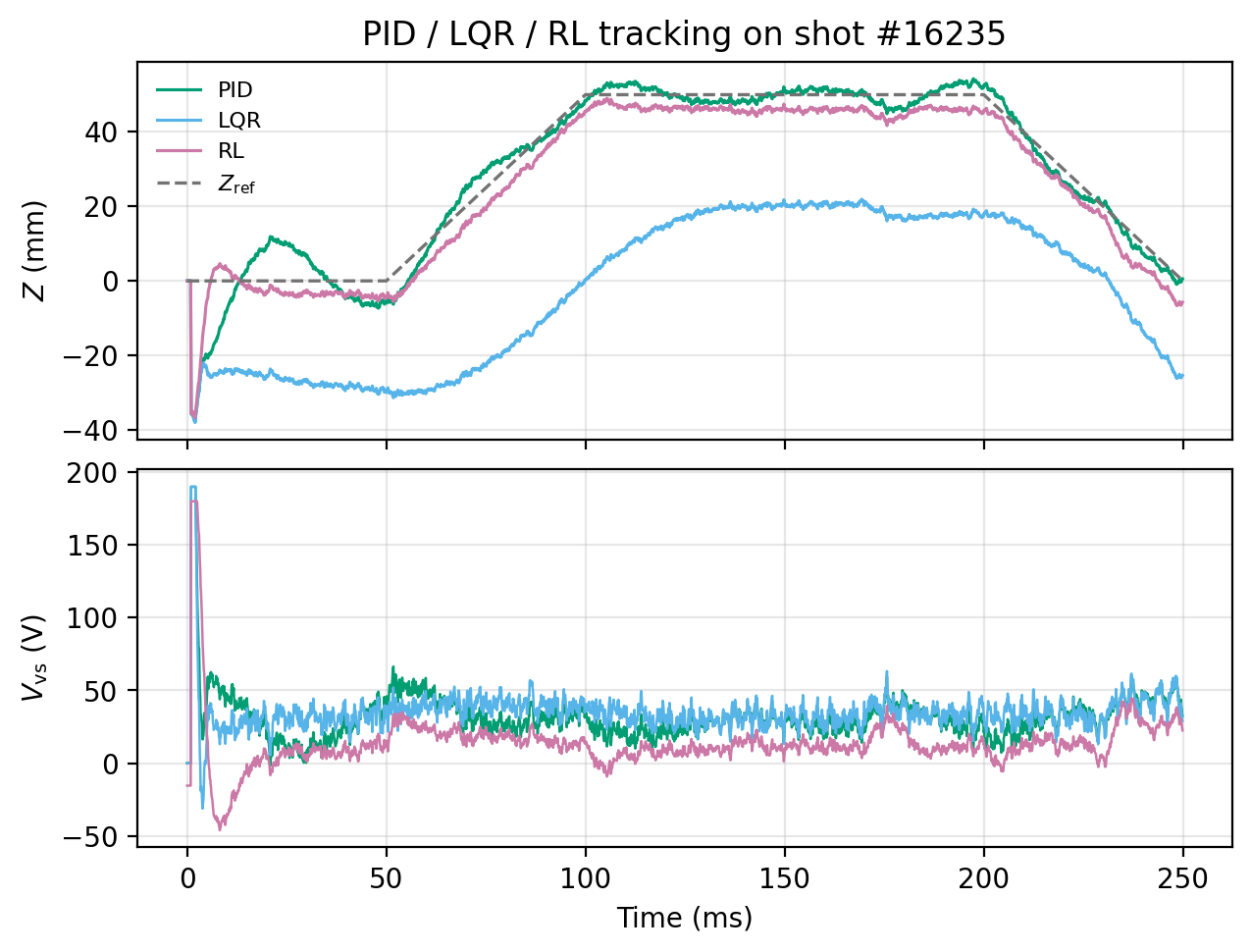}
  \caption{PID / LQR / RL on shot~\#16235
  (design/training on \#17567).
  Top: $Z$ and $Z_{\mathrm{ref}}$; bottom: $V_{\mathrm{vs}}$.}
  \label{fig:pid-lqr-rl-16235}
\end{figure}

\Cref{fig:pid-lqr-rl-16235} compares the three baseline laws of
\cref{sec:sim-setup} on shot~\#16235
(controllers fixed from \#17567; shared noise and voltage limits as in
\cref{sec:sim-env-verify}).
PID tracks the ramp--hold--return command with modest oscillation about
$Z_{\mathrm{ref}}$ and meets the settling band
($t_s\approx 77\,\mathrm{ms}$).
Pure LQR, by contrast, retains a large vertical offset and never enters the
$5\,\mathrm{mm}$ error band for a full dwell---consistent with missing integral
action under model--plant mismatch on a transfer shot.
The RL policy follows the commanded shape with a small residual undershoot
relative to PID but lower $V_{\mathrm{rms}}$
(\cref{tab:tracking-16235}).
Integral-augmented variants are taken up next.

\subsection{Integral compensation ablation}
\label{sec:sim-integral}

\begin{figure}[htbp]
  \centering
  \includegraphics[width=0.92\linewidth]{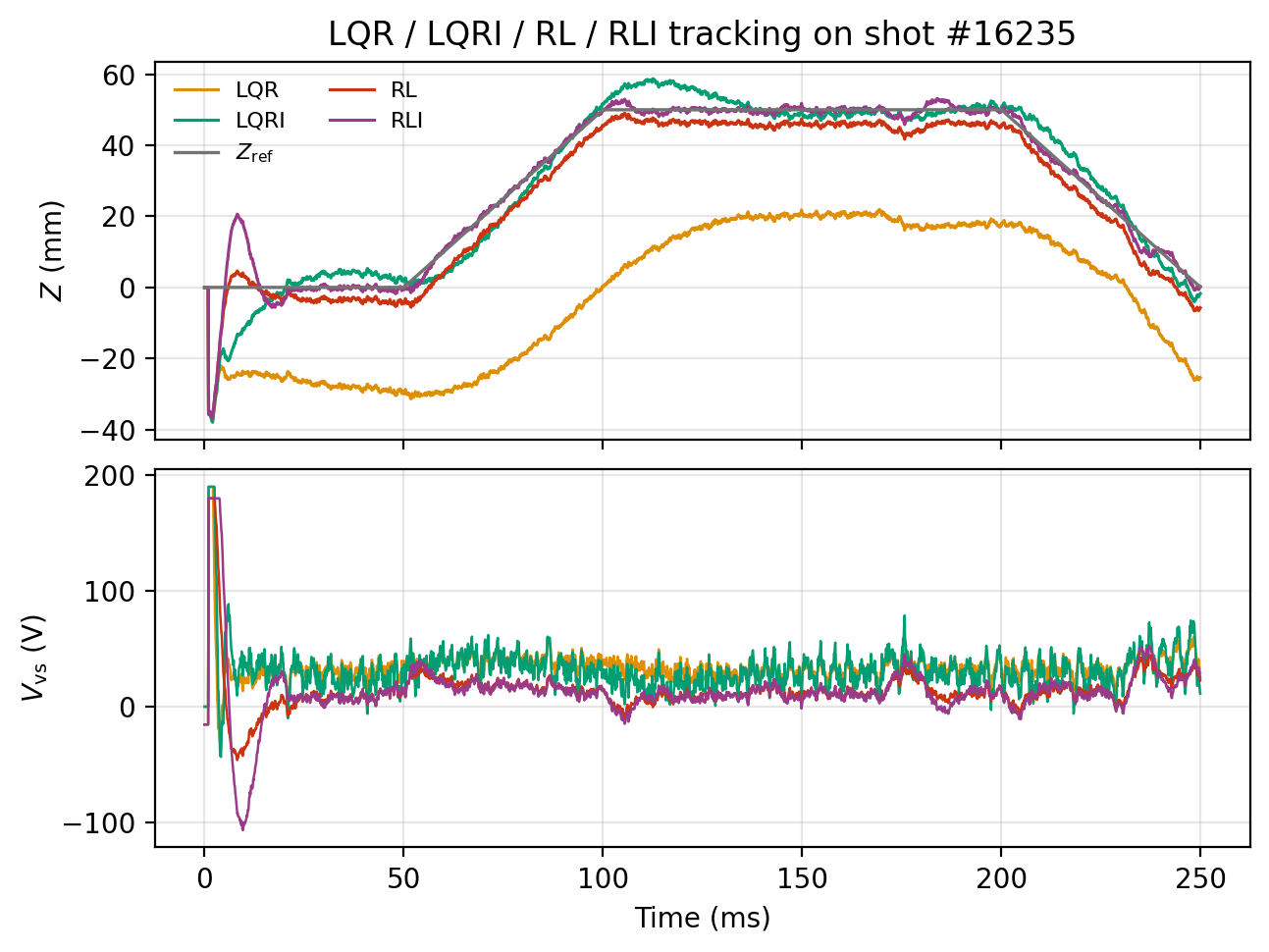}
  \caption{LQR / LQRI / RL / RLI on shot~\#16235
  (design/training on \#17567).
  Top: $Z$ and $Z_{\mathrm{ref}}$; bottom: $V_{\mathrm{vs}}$.}
  \label{fig:lqri-rli-16235}
\end{figure}

We next ablate the integral compensation of~\cref{sec:integral-comp}
on the same \#16235 case as~\cref{sec:sim-compare}:
LQR$\to$LQRI
($Q=\mathrm{diag}(10,\,10^{-3},\,3\times10^{5})$, $R_u=10^{-5}$)
and RL$\to$RLI, with plant, noise, and voltage limits unchanged.
\Cref{fig:lqri-rli-16235} overlays the four trajectories
(colour distinguishes the laws; all curves solid).
\Cref{tab:tracking-16235} collects all five controllers under one settling
rule ($t_s$ scored for $t\ge 10\,\mathrm{ms}$).

\begin{table}[t]
  \centering
  \caption{Closed-loop tracking indices on shot~\#16235
  (noise on; controllers from \#17567).
  Settling time $t_s$: $|e|\le 5\,\mathrm{mm}$ for $\ge 20\,\mathrm{ms}$,
  first attained for $t\ge 10\,\mathrm{ms}$ (\cref{eq:kpi-ts}).
  A dash means the criterion is not met.
  Baseline and integral-augmented rows are grouped for the ablation.}
  \label{tab:tracking-16235}
  \begin{tabular}{@{}lrrrr@{}}
    \toprule
    Controller & $t_s$ (ms) & ITAE ($10^{-5}\,\mathrm{m\cdot s^{2}}$)
      & OS (mm) & $V_{\mathrm{rms}}$ (V) \\
    \midrule
    PID & 77 & 6.67 & 37.9 & 35.7 \\
    LQR & --- & 96.07 & 51.2 & 37.7 \\
    RL & 10 & 14.14 & 36.8 & 24.4 \\
    \midrule
    LQRI & 79 & 9.20 & 37.9 & 36.8 \\
    RLI & 19 & 3.06 & 36.8 & 30.3 \\
    \bottomrule
  \end{tabular}
\end{table}

The integral add-on removes the large plateau offset of pure LQR: LQRI settles
($t_s\approx 79\,\mathrm{ms}$) and cuts ITAE by roughly an order of magnitude
relative to LQR, at nearly the same $V_{\mathrm{rms}}$.
For the primary RL path the tracking benefit is clearer in ITAE---RLI
attains the lowest value in the suite
($3.06\times10^{-5}\,\mathrm{m\cdot s^{2}}$) while following both ramps,
at a moderate increase in actuator effort relative to RL without $I$.
Thus integral compensation acts as a low-cost mismatch remedy on both the
model-based baseline and the RL primary controller---with different
realizations---rather than as a separate algorithmic family; subsequent
discussion therefore treats LQRI and RLI as the integral-augmented
counterparts of LQR and RL.

\subsection{Cross-shot check: RLI on four plants}
\label{sec:sim-adapt}

\begin{figure}[htbp]
  \centering
  \includegraphics[width=0.92\linewidth]{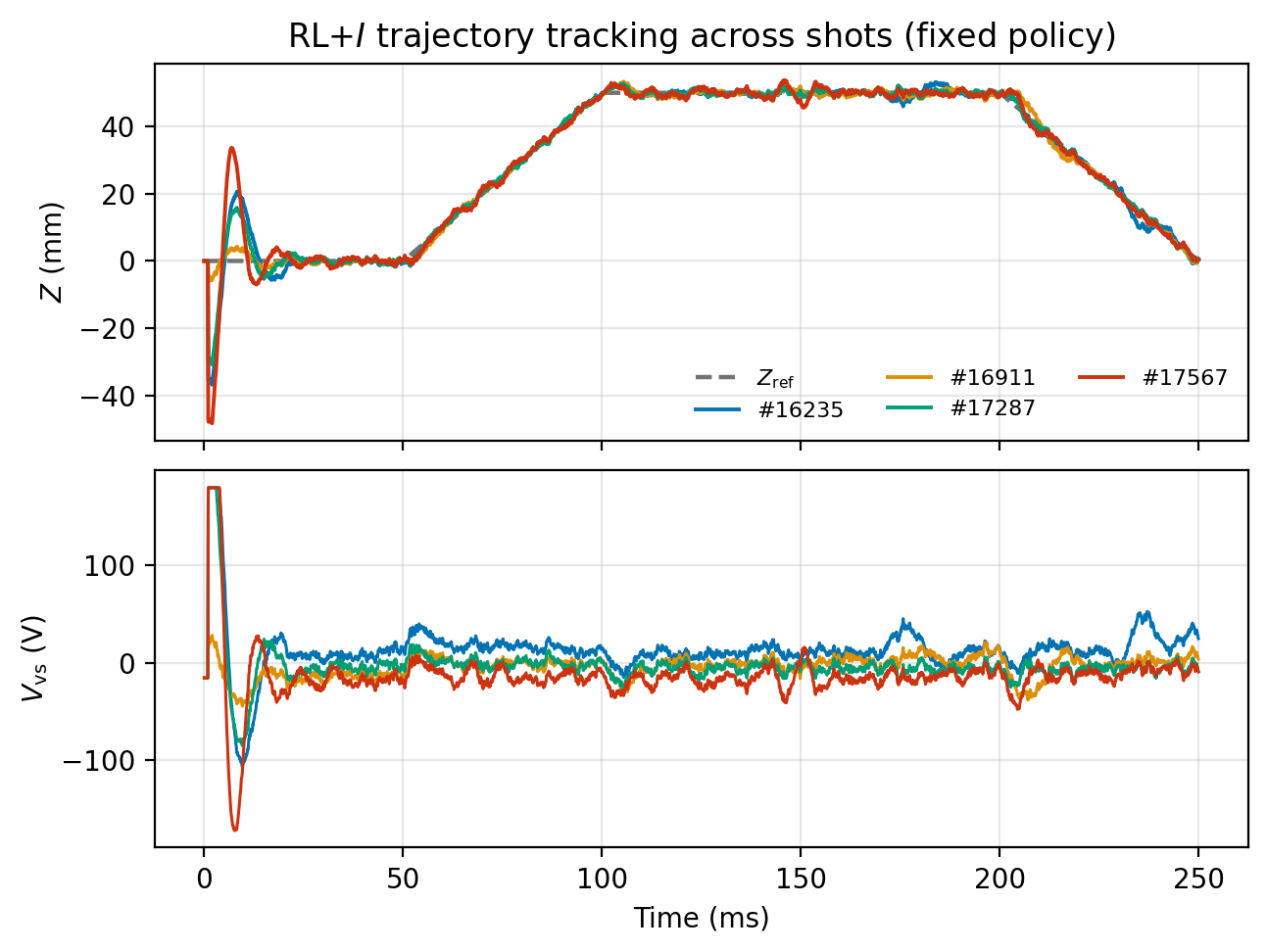}
  \caption{Fixed-policy RLI on four reconstructed plants under the same
  ramp--hold--return reference (trained on \#17567; no per-shot retuning).
  Top: $Z$ and $Z_{\mathrm{ref}}$; bottom: $V_{\mathrm{vs}}$.}
  \label{fig:rl-adapt-traj}
\end{figure}

The baseline and integral studies already probe transfer by design
(\#17567$\to$\#16235).
Here we keep the trained RLI policy fixed and apply the same
ramp--hold--return reference on four reconstructed plants
(\#16235, \#16911, \#17287, \#17567), without per-shot fine-tuning
(\cref{fig:rl-adapt-traj}, \cref{tab:rl-adapt-traj}).
The initial offset recovers on every plant
($t_s\approx 10$--$19\,\mathrm{ms}$ under the same rule as
\cref{eq:kpi-ts}), after which all four track the commanded excursion with
ITAE in a narrow band
($\approx 2.0$--$3.1\times10^{-5}\,\mathrm{m\cdot s^{2}}$).
Peak deviation and $V_{\mathrm{rms}}$ vary with the plant (largest OS on the
training shot \#17567; quietest actuation on \#16911), but vertical tracking
remains intact on every scenario.
Thus the integral-augmented primary controller transfers across this
four-shot suite in simulation; on-device multi-shot takeover is reported
in~\cref{sec:experiment}.

\begin{table}[t]
  \centering
  \caption{Fixed-policy RLI tracking indices on four plants
  (same $t_s$ rule as \cref{tab:tracking-16235}; ITAE/OS/$V_{\mathrm{rms}}$
  over the full window).}
  \label{tab:rl-adapt-traj}
  \begin{tabular}{@{}lrrrr@{}}
    \toprule
    Shot & $t_s$ (ms) & ITAE ($10^{-5}\,\mathrm{m\cdot s^{2}}$)
      & OS (mm) & $V_{\mathrm{rms}}$ (V) \\
    \midrule
    \#16235 & 19 & 3.06 & 36.8 & 30.3 \\
    \#16911 & 10 & 2.65 & 6.2 & 11.9 \\
    \#17287 & 15 & 2.01 & 30.8 & 23.5 \\
    \#17567 & 14 & 2.93 & 48.3 & 33.1 \\
    \bottomrule
  \end{tabular}
\end{table}

\begin{figure}[htbp]
  \centering
  \includegraphics[width=\linewidth]{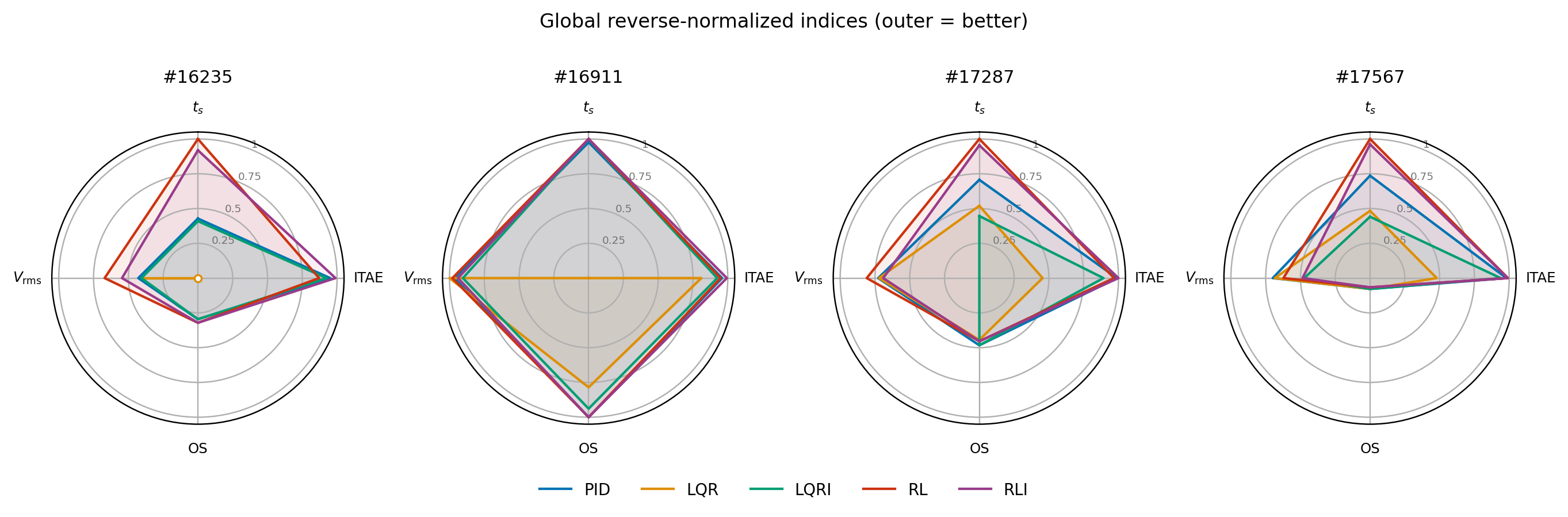}
  \caption{Cross-shot radar of PID / LQR / LQRI / RL / RLI
  (outer~$=$~better; one panel per shot).
  Axes: $t_s$, ITAE, OS, and $V_{\mathrm{rms}}$.
  Radial score: global reverse min--max~\eqref{eq:radar-score}.}
  \label{fig:sim-methods-radar}
\end{figure}

The same four-shot suite is summarized for all five controllers in the
radar of~\cref{fig:sim-methods-radar} (one panel per shot; axes $t_s$,
ITAE, OS, $V_{\mathrm{rms}}$).
Read the plot with a single rule first: \emph{outer~$=$~better}.
Under that reading, RLI is frequently among the outer envelopes, while pure
RL remains competitive---especially on settling and actuator effort---rather
than dominating every tracking axis (on the \#16235 transfer case, PID still
records a lower ITAE than RL without~$I$).
Pure LQR stays near the centre.
The radar therefore supports selecting an RL-centred path for on-device tests
on the basis of a favourable accuracy--effort trade-off across shots, not a
claim that RL outperforms PID on all tracking indices.

The radial coordinate is a score, not the raw index.
Each raw index of~\cref{sec:sim-setup} improves when smaller; to put them on
one scale we reverse min--max normalize \emph{globally} over all five
controllers and four shots,
\begin{equation}
  s
  =
  1
  -
  \frac{x-x_{\min}}{x_{\max}-x_{\min}},
  \qquad
  x_{\min}
  =
  \min_{c,k}\,x_{c,k},\quad
  x_{\max}
  =
  \max_{c,k}\,x_{c,k},
  \label{eq:radar-score}
\end{equation}
where $c$ runs over \{PID, LQR, LQRI, RL, RLI\} and $k$ over the four shots.
Thus $s=1$ maps to the outer ring (best sample of that index) and $s=0$ to
the centre (worst), which is why outer means better.
If $t_s$ is undefined (settling criterion not met), we assign $s=0$ for that
axis and exclude the undefined value from $x_{\min}$/$x_{\max}$.
Absolute values remain those in
\cref{tab:tracking-16235,tab:rl-adapt-traj} and the accompanying KPI table.

\subsection{Discussion}
\label{sec:sim-discuss}

Taken together, the simulation campaign supports the three-step narrative of
this section and the roles assigned in~\cref{sec:control_methods}.

\paragraph{Environment.}
Closed-loop verification against EXL-50U measurements
(\cref{sec:sim-env-verify}) indicates that the discharge-reconstructed LTV
plant, with the retained actuator limits and measurement chain, is a credible
testbed for controller comparison prior to deployment.

\paragraph{PID / LQR / RL baselines.}
On the \#17567$\to$\#16235 transfer case (\cref{sec:sim-compare}), the
engineering PID baseline tracks the ramp--hold--return command with the lowest
ITAE among the three laws.
Pure LQR exposes the cost of missing integral action under model--plant
mismatch (persistent plateau offset).
The primary RL policy follows the commanded shape with tracking accuracy
comparable to PID and consistently lower actuator effort
(\cref{fig:pid-lqr-rl-16235}, \cref{tab:tracking-16235}), even though its
ITAE on this transfer shot is not the lowest of the three.

\paragraph{Integral compensation as mismatch remedy.}
Reusing the same plant and reference (\cref{sec:sim-integral}), the integral
compensation of~\cref{sec:integral-comp} restores settling for LQR$\to$LQRI and
further improves tracking for RL$\to$RLI, which attains the best ITAE in
the suite at moderate $V_{\mathrm{rms}}$
(\cref{fig:lqri-rli-16235}, \cref{tab:tracking-16235}).
The add-on is therefore best read as a low-cost compensation for residual
mismatch on both the model-based and learning paths---not as a fourth
controller family, and not as a single shared implementation.

\paragraph{Toward on-device validation.}
Beyond the \#17567$\to$\#16235 protocol, the fixed RLI trajectories
(\cref{fig:rl-adapt-traj}, \cref{tab:rl-adapt-traj}) and the four-shot radar
(\cref{fig:sim-methods-radar}) jointly support transferring an RL-centred
path across reconstructed plants (outer~$=$~better), with the clearest
gains appearing once integral compensation is included (RLI).
\Cref{sec:experiment} assesses that primary path and the LQR baseline under
real EXL-50U discharge conditions.

\section{On-device experiments}
\label{sec:experiment}

The simulation study of~\cref{sec:simulation} selects an RL primary path
and an LQR model-based baseline for deployment.
This section reports closed-loop vertical-position control on EXL-50U.
The RL policy has been taken over on more than ten discharges with stable
regulation inside the commanded control window.
Here we document representative time traces---RL takeover on \#20177 and LQR
control on \#19727 (\cref{fig:exp-lqr-rl})---and a multi-shot comparison of
RL against the operational PID baseline using the tracking indices of
\cref{sec:sim-setup} (\cref{fig:exp-rl-pid-box}).

\subsection{Experimental setup}
\label{sec:exp-setup}

Controllers follow the designs of~\cref{sec:control_methods}, with gains /
policies frozen from the simulation synthesis (no pulse-wise retuning in the
windows shown).
Vertical position $Z$ and VS voltage $V_{\mathrm{VS}}$ are taken from the
EXL-50U MDS archive; plasma current $I_p$ is shown as a discharge-phase
context.
Each controller is active only inside a prescribed takeover window (shaded in
\cref{fig:exp-lqr-rl}); outside that window the plant reverts to the
operational PCS path.
Within the active window we quote two simple regulation indices on $Z$:
the standard deviation $\sigma_Z$ and the peak-to-peak excursion
$\Delta Z_{\mathrm{p\text{-}p}}$, both in centimetres.
Because the RL and LQR pulses differ in $I_p$ waveform and in the length of
the controlled interval, the indices are interpreted as descriptive window
statistics rather than a ranked head-to-head score.

\subsection{Representative RL and LQR pulses}
\label{sec:exp-represent}

\begin{figure}[htbp]
  \centering
  \includegraphics[width=\linewidth]{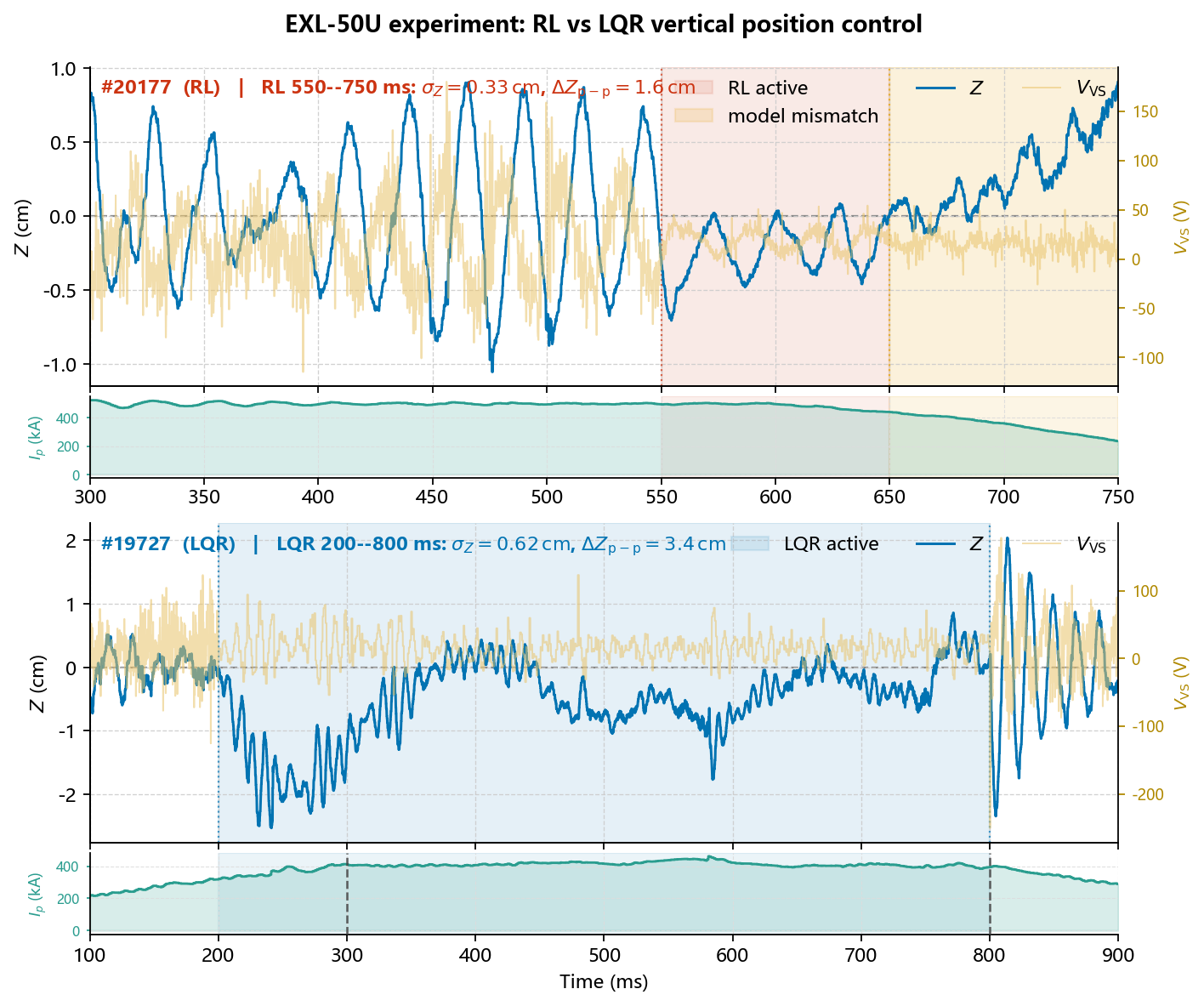}
  \caption{EXL-50U closed-loop vertical control: RL takeover on shot~\#20177
  (top) and LQR on shot~\#19727 (bottom).
  Shaded bands mark the active controller window; on \#20177 the orange
  $650$--$750\,\mathrm{ms}$ band marks the $I_p$ ramp-down
  (\emph{model mismatch}), where tracking degrades relative to the preceding
  flattop segment.
  Annotated $\sigma_Z$ and $\Delta Z_{\mathrm{p\text{-}p}}$ are computed on
  the window indicated in each panel title
  ($550$--$750\,\mathrm{ms}$ for RL; full LQR active window for \#19727).
  Lower strips show $I_p$ for discharge-phase context.}
  \label{fig:exp-lqr-rl}
\end{figure}

\Cref{fig:exp-lqr-rl} contrasts the two pulses.
On shot~\#20177 the RL controller is active from $550$ to $750\,\mathrm{ms}$.
Over the first $100\,\mathrm{ms}$ of takeover ($550$--$650\,\mathrm{ms}$),
while $I_p$ remains near flattop, $Z$ stays close to the null.
In the subsequent $650$--$750\,\mathrm{ms}$ band marked
\emph{model mismatch} on the figure, $I_p$ ramps down and the regulation
quality visibly degrades: $Z$ drifts away from zero and the oscillation
grows, even though the same frozen policy remains in command.
The annotated $\sigma_Z$ and $\Delta Z_{\mathrm{p\text{-}p}}$ are scored
over the full $550$--$750\,\mathrm{ms}$ takeover and therefore mix the
quieter flattop segment with this ramp-down interval.
Large $Z$ motion after the takeover ends is outside the claimed control
window.

On shot~\#19727 the LQR controller is active over a longer interval
($200$--$800\,\mathrm{ms}$).
An initial transient after engagement is followed by regulation about $Z=0$
with $\sigma_Z\approx 0.62\,\mathrm{cm}$ and
$\Delta Z_{\mathrm{p\text{-}p}}\approx 3.4\,\mathrm{cm}$ over the scored
LQR window, while $I_p$ evolves through ramp-up and flattop.
As with the RL pulse, departure from the null after the shaded window is not
attributed to the tested law.

\subsection{Multi-shot PID vs RL comparison}
\label{sec:exp-rl-pid}

\begin{figure}[htbp]
  \centering
  \includegraphics[width=\linewidth]{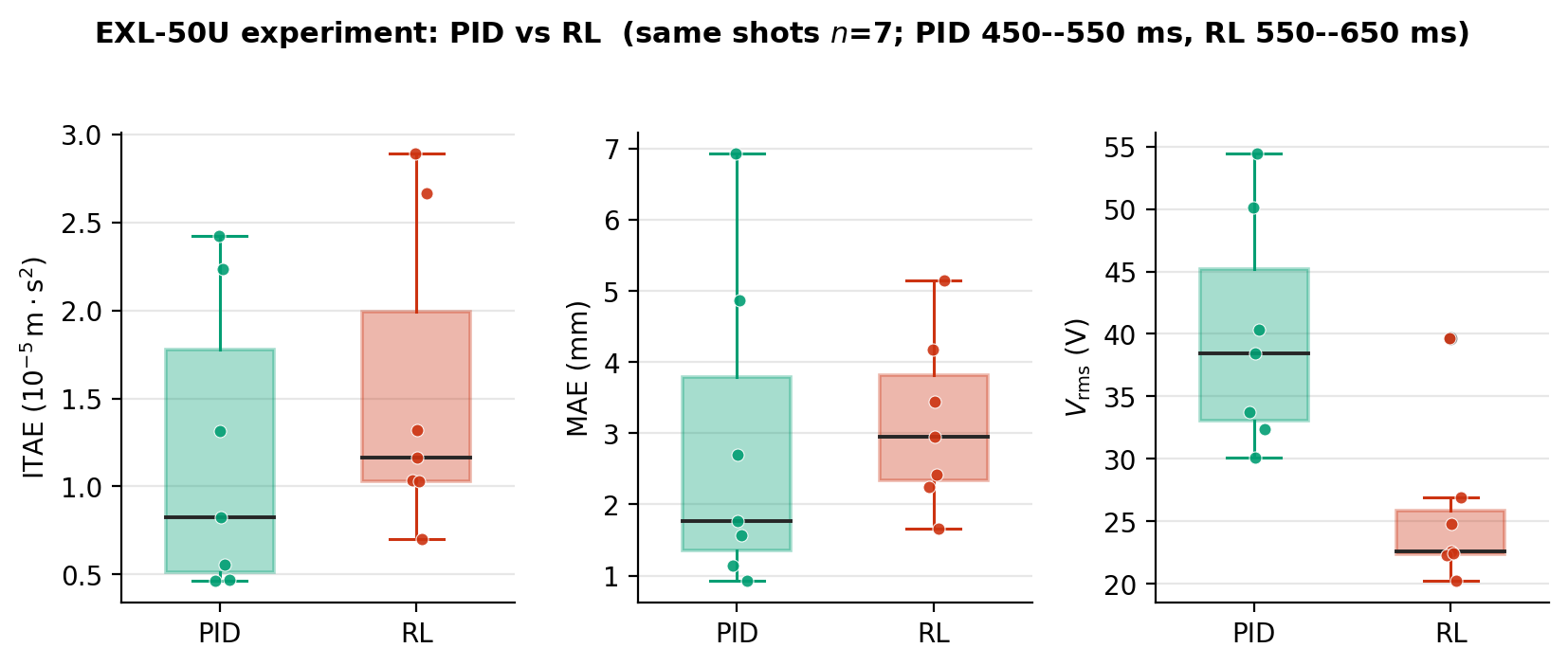}
  \caption{On-device PID vs RL box plots on the same seven shots:
  ITAE, MAE (\cref{eq:exp-mae}), and $V_{\mathrm{rms}}$ ($Z_{\mathrm{ref}}=0$).
  Adjacent windows of $T=100\,\mathrm{ms}$:
  PID $450$--$550\,\mathrm{ms}$; RL $550$--$650\,\mathrm{ms}$.
  Points mark individual pulses.}
  \label{fig:exp-rl-pid-box}
\end{figure}

To compare the RL primary path with the operational PID baseline across
pulses, we evaluate ITAE and $V_{\mathrm{rms}}$ from~\cref{sec:sim-setup}
together with the mean absolute error
\begin{equation}
  \mathrm{MAE}
  =\frac{1}{T}\int_{0}^{T} \bigl|e(t)\bigr|\,\mathrm{d}t
  =\frac{\mathrm{IAE}}{T}
  \label{eq:exp-mae}
\end{equation}
(here $Z_{\mathrm{ref}}=0$) on the \emph{same} seven RL-takeover shots
(\#20150, \#20151, \#20173, \#20174, \#20175, \#20177, \#20179), using
adjacent equal-length windows $T=100\,\mathrm{ms}$: operational PID on
$450$--$550\,\mathrm{ms}$ (immediately before takeover) and RL on
$550$--$650\,\mathrm{ms}$ (active window), so that $I_p$ and the equilibrium
phase remain comparable.
Time in ITAE is measured from the start of each scored window.
\Cref{fig:exp-rl-pid-box} summarizes the distributions.
MAE remains in the few-millimetre range for both controllers on most pulses
(typically $\sim 1$--$5\,\mathrm{mm}$), and ITAE is of the same order;
$V_{\mathrm{rms}}$ is lower under RL on every shot in this set.

\subsection{Discussion}
\label{sec:exp-discuss}

The representative pulses confirm that both the RL primary controller and the
LQR baseline can regulate vertical position inside a commanded takeover
window, with centimetre-scale $\sigma_Z$ on \#20177 consistent with the
divertor-era target of~\cref{sec:introduction}.
At the same time, the orange band on \#20177
(\cref{fig:exp-lqr-rl}) shows that a frozen policy is less effective once
$I_p$ enters the ramp-down: the associated plant change acts as a clear
model--plant mismatch and the vertical tracking quality drops relative to the
flattop segment.
On the same-shot, adjacent-window comparison
(\cref{fig:exp-rl-pid-box}), which scores only the flattop-adjacent
$550$--$650\,\mathrm{ms}$ interval under comparable discharge conditions
(not a simultaneous A/B test), RL achieves millimetre-scale tracking
accuracy comparable to operational PID while consistently reducing VS
effort ($V_{\mathrm{rms}}$) on every pulse examined.
Together with the broader RL-takeover campaign ($>$10 shots with stable
regulation in the controlled window), these results support deploying RL as
the primary on-device path alongside the LQR baseline, as motivated
by~\cref{sec:simulation}.
Improving regulation through the current ramp-down will likely require
stronger D-layer remedies---more robust synthesis and/or online or
pulse-wise parameter identification---beyond the integral compensation used
here.

\section{Conclusion}
\label{sec:conclusion}

This work has addressed vertical stability control for EXL-50U, motivated by a
high VDE-linked disruption fraction in recent operations and by the gap between
millimetre-level limiter control and centimetre-scale divertor oscillations
under the present PID loop.
A reinforcement-learning policy is developed as the primary controller, with
classical PID and LQR retained as engineering and model-based baselines, and
with integral compensation as a low-cost remedy for model--plant /
training--deployment mismatch (state augmentation for LQR; external integral
for RL).

A high-fidelity, discharge-reconstructed simulation environment supports
synthesis, training, and ablation
(\cref{sec:simulation}): PID tracks the commanded excursion while pure LQR
retains a plateau offset and RL remains competitive at lower actuator effort;
integral augmentation restores LQR$\to$LQRI settling and further improves
RL$\to$RLI, which then transfers across four reconstructed plants under a
fixed policy.

On-device tests (\cref{sec:experiment}) confirm deployability.
Representative pulses \#20177 (RL) and \#19727 (LQR) regulate $Z$ inside
commanded takeover windows, and across more than ten RL-takeover discharges
the policy remained stable in the controlled interval, with flattop accuracy
at the sub-centimetre level (down to a few millimetres in favourable cases).
On seven of these shots, adjacent equal-length windows within the same
discharge (PID $450$--$550\,\mathrm{ms}$; RL $550$--$650\,\mathrm{ms}$)
show millimetre-scale MAE for both controllers and systematically lower
$V_{\mathrm{rms}}$ under RL, indicating that RL achieves tracking accuracy
comparable to the operational PID baseline while consistently reducing
actuator effort under comparable discharge conditions.

The integral compensation used here remains a coarse entry in that
sim2real~/~mismatch menu.
In particular, the $I_p$ ramp-down segment of the on-device RL pulse
(\cref{fig:exp-lqr-rl}) already shows a clear drop in tracking quality under
a frozen controller, indicating that more robust designs and/or
parameter-identification-based adaptation will be needed to sustain
performance through such phase changes.
More structured remedies---real-time or pulse-wise system identification,
state estimation, and robust synthesis in the spirit of $H_\infty$ / related
methods---are therefore left for future development, together with a broader
matched LQR catalogue and careful treatment of delay, saturation, and
growth-rate variation as the on-device campaign is expanded.


\FloatBarrier
\clearpage
\printbibliography

\end{document}